# Exceptionally high phonon-limited carrier mobility in BX (X = P, As, Sb) monolayers

Shiru Song,[†] Shixu Liu,[†] Yuting Sun,[†] Ji-Hui Yang[†‡]* and Xin-Gao Gong[†‡]

[†]Key Laboratory of Computational Physical Sciences (Ministry of Education), Institute of Computational Physics, Fudan University, Shanghai 200433, China
[‡]Shanghai Qi Zhi Institute, Shanghai 200230, China

**ABSTRACT:**

　Ideal two-dimensional (2D) semiconductors with high mobility comparable to three-dimensional (3D) Si or GaAs are still lacking, hindering the development of high-performance 2D devices. Here in this work, using first-principles calculations and considering all the electron-phonon couplings, we show that monolayer BX (X = P, As, Sb) with honeycomb lattices have intrinsic phonon-limited carrier mobility reaching record-high values of 1200~14000 cm$^2$V$^{-1}$s$^{-1}$ at room temperature. Despite being polar and the band edges located at the K point with multiple valleys, these three systems unusually have small carrier scattering rates. Detailed analysis shows that, both the intravalley scattering and the intervalley scattering between two equivalent K points are weak, which can be understood from the large mismatch between the electron bands and phonon spectrum and suppressed electron-phonon coupling strength. Furthermore, we reveal the general trend of mobility increase from BP to BAs and to BSb and conclude that: smaller effective masses, larger sound velocities, higher optical phonon energies, heavy atomic masses, and out-of-plane orbitals tend to result in small match between the electron and phonon bands, small electron-phonon coupling strengths, and thus high mobility. Our work demonstrates that 2D semiconductors can achieve comparable carrier mobility to 3D GaAs, thus opening doors to 2D high-performance electronic devices.

**KEYWORDS:** high intrinsic mobility, two-dimensional, boride honeycomb lattice, first-principles calculation, scattering rate

## INTRODUCTION

　The next generation of electronic devices with high performance and low-power consumption has made two-dimensional (2D) semiconductors attractive ever since the first exfoliation of graphene with atomic thickness.[1–11] In general, an ideal 2D semiconductor for electronic applications should have high mobility at room temperature to ensure fast response. Current widely studied 2D semiconductors, including transition metal dichalcogenides (TMDs), black phosphorus, and InSe, all have typical carrier mobility of less than several hundred $cm^2V^{-1}s^{-1}$, which is much smaller than the electron mobility of three-dimensional (3D) silicon and GaAs.[12–19] Therefore, to achieve comparable performance of devices based on 3D semiconductors, it's necessary to find 2D semiconductors with high mobility, which has been one of the most important issues in the current 2D community.

So far, there are over one thousand 2D materials proposed.[20–28] The question is, which or what kind of 2D semiconductor is likely to have high mobility? In fact, many 2D semiconductors have been predicted to have high mobility[29–36] based on the deformation potential approximation (DPA),[37] which just considers the coupling between longitudinal acoustic (LA) phonon modes and electrons. However, recent calculations by considering all the electron-phonon couplings (EPC),[15,18,38,39] which are also computationally expensive of course, show that DPA generally overestimates mobility too much and leads to unreliable results. This is because the electron scattering by longitudinal optical (LO) phonon modes is also very important, especially for polar systems as reported by Cheng et al.[38] that the carrier mobility of most TMDs is mainly limited by LO phonons. In their opinion, 2D nonpolar semiconductors are better potential candidates to have high mobility. They further reported 2D antimony has high hole mobility of 1330 $cm^2V^{-1}s^{-1}$ owing to its band structure with a steep and deep valley located at the Γ point, which suppresses intervalley scattering.[39] However, the electron mobilities of 2D group V monolayers are still very low. As there are only a few 2D nonpolar semiconductors with Γ valleys, one would wonder whether ideal 2D systems with high mobility can be found or not.

Our recent work showed that multiple valleys at the non-Γ points actually do NOT necessarily lead to strong intervalley scattering.[40] We further proposed a matching function to capture both the intervalley and intravalley scatterings and find that the matching function has no well-defined relations with the system polarity and the number of valleys at the band edges. As a result, 2D semiconductors with high mobility should not be limited to nonpolar and single Γ-valley systems but can be reasonably expanded by screening the matching functions. Motivated by this, here in this work, using first-principles calculations and considering all the electron-phonon couplings, we find that monolayer BX (X = P, As, Sb) with honeycomb lattices have exceptionally high both electron and hole mobility at room temperature (1200~14000 $cm^2V^{-1}s^{-1}$), although they are polar systems, and their band edges are located at the K point with multiple valleys. Especially, monolayer BSb has intrinsic electron and hole mobility of 8956.3 cm$^2$V$^{-1}$s$^{-1}$ and 14033.4 cm$^2$V$^{-1}$s$^{-1}$, respectively, which are the highest among known 2D materials. We attribute the high mobility to the weak intravalley scattering and the weak intervalley scattering between two equivalent K points due to the large mismatch between the electron bands and phonon spectrum and suppressed electron-phonon coupling strengths. Our findings demonstrate that 2D semiconductors do have potentials to achieve comparably high electronic performance relative to 3D systems.

**CALCULATION METHODS**

We calculate phonon-limited intrinsic carrier mobility using the ab initio Boltzmann transport equation in the framework of self-energy relaxation time approximation.[19] The mobility is given by

$$\mu_{\alpha\beta} = \frac{-1}{V_{uc}n_c}\sum_n \int \frac{d^3k}{\Omega_{BZ}} \frac{\partial f^0_{n\mathbf{k}}}{\partial \varepsilon_{n\mathbf{k}}} v_{n\mathbf{k},\alpha} v_{n\mathbf{k},\beta} \tau_{n\mathbf{k}} \quad (1),$$

where $n_c$ is the carrier concentration, $V_{uc}$ is the volume of unit cell, $\Omega_{BZ}$ denotes the volume of the first Brillouin zone, $v_{n\mathbf{k},\alpha}$ is the band velocity for the Kohn-Sham state $|n\mathbf{k}\rangle$, and $f^0_{n\mathbf{k}}$ is the Fermi-Dirac equilibrium occupation at a given temperature, i.e., at room temperature. Here, $\alpha$, $\beta$ run over the three Cartesian directions, the summation is over the band index $n$, and the integral is over the electron wavevectors **k** in the first Brillouin zone. The scattering rate, given by the reciprocal of the carrier scattering lifetime $\tau_{n\mathbf{k}}$, is expressed as:

$$\tau_{n\mathbf{k}}^{-1} = \frac{2\pi}{\hbar}\sum_{m,\nu} \int \frac{d^3q}{\Omega_{BZ}} |g_{mn\nu}(\mathbf{k},\mathbf{q})|^2 \big[(1 - f^0_{m\mathbf{k}+\mathbf{q}} + n_{\mathbf{q}\nu})\delta(\varepsilon_{n\mathbf{k}} - \varepsilon_{m\mathbf{k}+\mathbf{q}} - \hbar\omega_{\mathbf{q}\nu}) +$$
$$(f^0_{m\mathbf{k}+\mathbf{q}} + n_{\mathbf{q}\nu})\delta(\varepsilon_{n\mathbf{k}} - \varepsilon_{m\mathbf{k}+\mathbf{q}} + \hbar\omega_{\mathbf{q}\nu})\big] \quad (2),$$

where **q** is a phonon wavevector, $n_{\mathbf{q}\nu}$ is the Bose-Einstein distribution and $\omega_{\mathbf{q}\nu}$ is phonon frequency. The summation runs over the electron band index $m$ and phonon branch index $\nu$. $g_{mn\nu}(\mathbf{k},\mathbf{q})$ is the electron-phonon matrix element, which evaluates the scattering amplitude from an initial state $n\mathbf{k}$ to a final state $m\mathbf{k}+\mathbf{q}$ via the emission or absorption of a phonon with indices $\mathbf{q}\nu$. The energy conservation during the scattering process is described by the Dirac delta functions. We use the EPW code[41,42] based on the first-principles Quantum Espresso Package[43,44] to calculate the electron-phonon couplings and carrier mobility. More parameters can be found in the Supporting Information.

To understand the scattering mechanisms, we decompose the scattering rate into phonon-energy-resolved contributions as[45]:

$$\tau_{n\mathbf{k}}^{-1} = \int d\omega\, \partial \tau_{n\mathbf{k}}^{-1}/\partial \omega \quad (3).$$

Here,

$$\partial \tau_{n\mathbf{k}}^{-1}/\partial \omega = \frac{2\pi}{\hbar}\sum_{m,\nu}\int \frac{d^3q}{\Omega_{BZ}}|g_{mn\nu}(\mathbf{k},\mathbf{q})|^2\big[(1 - f_{m\mathbf{k+q}}^0 + n_{\mathbf{q}\nu})\delta(\varepsilon_{n\mathbf{k}} - \varepsilon_{m\mathbf{k+q}} - \hbar\omega_{\mathbf{q}\nu}) + (f_{m\mathbf{k+q}}^0 + n_{\mathbf{q}\nu})\delta(\varepsilon_{n\mathbf{k}} - \varepsilon_{m\mathbf{k+q}} + \hbar\omega_{\mathbf{q}\nu})\big]\delta(\omega - \omega_{\mathbf{q}\nu}) \quad (4),$$

which reflects the density of scattering rate of an electron at an initial state $|n\mathbf{k}\rangle$ via the emission or absorption of a phonon with an energy of $\hbar\omega$. Further, we decompose $\partial \tau_{n\mathbf{k}}^{-1}/\partial \omega$ into two terms as the followings:

$$\partial \tau_{n\mathbf{k}}^{-1}/\partial \omega = 2\pi|g_{n\mathbf{k}}^*(\omega)|^2 F(\omega) \quad (5).$$

Here,

$$F(\omega) = \frac{1}{\hbar}\sum_{m,\nu}\int \frac{d^3q}{\Omega_{BZ}}\big[(1 - f_{m\mathbf{k+q}}^0 + n_{\mathbf{q}\nu})\delta(\varepsilon_{n\mathbf{k}} - \varepsilon_{m\mathbf{k+q}} - \hbar\omega_{\mathbf{q}\nu}) + (f_{m\mathbf{k+q}}^0 + n_{\mathbf{q}\nu})\delta(\varepsilon_{n\mathbf{k}} - \varepsilon_{m\mathbf{k+q}} + \hbar\omega_{\mathbf{q}\nu})\big]\delta(\omega - \omega_{\mathbf{q}\nu}) \quad (6),$$

which is phonon-energy resolved matching function between electronic band structure and phonon spectrum proposed in our recent work.[40] Another term $g_{n\mathbf{k}}^*(\omega)$ can be seen as phonon-energy resolved average EPC strength.

**RESULTS AND DISCUSSIONS**

By examining the electronic band structures of group III-V monolayers,[46] we find that monolayer nitrides have rather flat bands while the bands in the monolayer borides are rather dispersed with band edges located at the K point with multiple valleys. Figure 1a show the band structure of BX using BSb as an example. We see that, they have direct band gaps with the band edges located at the K point, in agreement with previous studies.[46] Note that, BX has multiple electron and hole valleys (see Fig. 1b), similar to monolayer MoS$_2$.[40] What is different from MoS$_2$ is that, the valance band maximum (VBM) of BX is mainly composed of X $p_z$ orbital and the conduction band minimum (CBM) is dominated by B $p_z$ orbital (see Figs. 1c and 1d). Since the $p_z$ orbitals are more delocalized than Mo $d$ orbitals, the bands in BX are more dispersed than those in MoS$_2$. Accordingly, the (average) effective carrier masses in BX are much smaller, i.e., 0.07 $m_e$ for electron and 0.07 $m_e$ for hole in BSb compared to 0.45 $m_e$ and 0.56 $m_e$ for electron and hole in MoS$_2$, respectively.

The calculated mobility results are listed in the Table 1. While InN has low mobility as expected, the borides have unusually high phonon-limited electron and hole mobility, all of which are larger than 1000 cm$^2$V$^{-1}$s$^{-1}$. Note that, our calculated carrier mobility is much smaller than the DPA predictions.[46] This is because, the DPA predictions just considered the carrier scatterings caused by the LA phonons. But actually, the scatterings by other phonons also have significant or even more important effects. As seen in Fig. 2a, there are six phonon branches in BX. The LA and transverse acoustic (TA) modes correspond to the translational vibrations. The LO mode is mainly the vibration of boron atoms and therefore, it has relatively high frequency due to the small atomic mass of boron. Besides, we also find that the LA mode at the K point (KLA) plays a very important role which corresponds to the vibration of Sb atoms. For BP, we find that both the hole and electron mobilities are limited nearly equally by TA and LA modes (see Figs. 2b and 2c as well as Tables S1 and S2). For BAs, both the electron and hole scatterings are more dominated by the TA modes rather than the LA modes. Besides, the LO mode and the LA mode at the K point (KLA) also cause significant scattering for electrons (see Fig. 2c and Table S1). For BSb holes, the LA and LO modes cause similar scattering rates, as seen in Fig. 2b and Table S2. For BSb electrons, the mobility is limited together by the LA, KLA and LO modes. In a word, DPA fails to evaluate the carrier mobility in these three systems and cannot give accurate results.

From BP to BAs and to BSb, the electron and hole mobilities gradually increase. Especially, monolayer BSb has electron mobility of 8956.3 cm$^2$V$^{-1}$s$^{-1}$, comparable to that of 3D GaAs. The hole mobility is 14033.4 cm$^2$V$^{-1}$s$^{-1}$, much larger than any known 2D semiconductors. With such high mobility, these three 2D systems are expected to offer great promise for high-performance 2D electronic devices comparable to current 3D systems.

Next, we try to understand why the carrier mobility in BX monolayers is so high despite they are polar and have multiple valleys. In the following discussions, we mainly use BSb monolayer as an example. While the small effective mass contributes to high mobility as expected, the scattering rate is another important factor. Note that, the effective electron mass of BSb is about one sixth of that of MoS$_2$ monolayer but the mobility is about 40 times larger. We find that BSb has a rather small scattering rate, i.e., about 0.6 THz in BSb versus about 8 THz in MoS$_2$ for the CBM states (see Fig. 3).

To understand why the scattering rate is so small in BSb monolayer, we consider the phonon-energy-resolved scattering rates, the matching function, and the average EPC strength as defined in the calculation methods. As seen in Figs. 3a and 3b, both the electron and hole scattering rates mainly have four contributions, as indicated by the four peaks. The first and second peaks are mainly contributed by the low-energy phonons (0–100 cm$^{-1}$ or 0–12 meV) corresponding to the intravalley scattering around the band edges, as seen from the phonon-branch-resolved scattering rates in Figs. 4a and 4b showing that the phonons involved in the scattering process mainly have very small momentum. The third peak, which is contributed by phonons around 200 cm$^{-1}$, comes from the intervalley scattering, as the contributions are mainly from phonons with momentum **q**=**K** (see Figs. 4a and 4b). The fourth peak distributed around 700 cm$^{-1}$ has both intra- and inter-valley contributions. This is reasonable as carriers can be scattered to larger regions with the assistance of high-energy phonons (~80 meV).

Note that, both the inter- and intra-valley scatterings in BSb are actually very small compared to known 2D systems. Because the band edges in BSb have phonon-energy-resolved scattering distributions similar to the CBM valley of MoS$_2$ monolayer, we make some comparable studies in the followings. Compared to the phonon-energy-resolved scattering rates for electrons in monolayer MoS$_2$,[40] we find that the smaller scattering rates in BSb are mainly owing to the lower scatterings in both the region of low-energy phonons with near zero momentum and the high-energy-phonon region (the fourth peak). First, in the low-energy-phonon region, we find that both the matching function $F(\omega)$ and the average EPC are much smaller in BSb than in MoS$_2$, leading to the much smaller intravalley scatterings. To reveal the origins, we consider one isotropic and parabolic electronic band with an effective mass of $m^*$ and one isotropic acoustic phonon branch with a sound velocity of $v$. In this case, we derive an analytical form of $F(\omega)$ which satisfies (see the Supporting Information):

$$F(\omega) \propto \frac{k_B T}{v^2} \quad (7),$$

where $k_B$ is the Boltzmann constant and T is the temperature. In BSb, we find that $F(\omega)$ is mainly contributed by the LA phonon mode with a sound velocity of about 1990 ms$^{-1}$ (see Table 2). On the other hand, $F(\omega)$ in MoS$_2$ is dominated by the TA phonon mode with a sound velocity of about 1230 ms$^{-1}$ (see Table 2). Therefore, $F(\omega)$ in BSb is about one sixth of that of MoS$_2$ considering the spin effect (Note that, spin-orbital coupling effect is considered in BSb but not in MoS$_2$). For the EPC, we find that this term is proportional to $\sqrt{\frac{m^*}{M}}$ (here $M$ is the effective atomic mass, see the Supporting Information) for scatterings caused by acoustic phonons. As BSb has a larger $M$, it has weaker EPC than MoS$_2$.

In the high-phonon-energy region (700 cm$^{-1}$ for BSb and 400 cm$^{-1}$ for MoS$_2$), we find that the smaller scattering rates in BSb are mainly caused by the smaller matching function $F(\omega)$ in BSb. To understand this, we consider one isotropic and parabolic electronic band with an effective mass of $m^*$ and one flat optical phonon branch with a constant frequency of $\omega_o$. In this case, we find that:

$$F(\omega) \propto m^*[\frac{1}{\exp(\frac{\hbar\omega_o}{k_B T})+1}+\frac{1}{\exp(\frac{\hbar\omega_o}{k_B T})-1}] \quad (8).$$

Because BSb has a smaller $m^*$ and larger $\omega_o$, it has a much smaller $F(\omega)$, leading to the smaller scattering rates in the high-energy-phonon regions.

One similar point in both BSb and MoS$_2$ monolayers is that the intervalley scattering between two equivalent $K$ points assisted by phonons with the momentum **q**=**K** is unexpectedly small, which can be attributed to two aspects. On one hand, the EPC strength, which is proportional to 1/q for optical phonons,[47,48] is the smallest at **q**=**K**. On the other hand, when the band edge states at the $K$ point are composed of out-of-plane orbitals, the carriers feel very small potential changes induced by atomic vibrations, leading to small EPC. Note that, the hole has weaker intervalley scatterings than the electron (see Fig. 3). This is because the VBM is composed of Sb $p_z$ orbitals and the intervalley scattering is mainly contributed by the phonon modes with the displacement of Sb. Therefore, the hole feels smaller potential changes as it moves with Sb, leading to smaller EPC compared to the case of the CBM electron located at boron atoms (see Figs. 3e and 3f).

Finally, we try to understand the increasing trend of carrier mobility from BP to BAs and to BSb monolayers. As shown in the Supporting Information, these three systems generally have very similar band structures, phonon spectra as well as scattering distributions. Note that, the intervalley scatterings in these systems are all weak (see the above discussions), especially for BP and BAs in which the scattering rates are mainly contributed by the intravalley scatterings caused by the low-energy phonons. From the above analysis, we already know that in the region of low-energy phonons, the matching function $F(\omega)$ is proportional to $1/v^2$ and the EPC is proportional to $\sqrt{\frac{m^*}{M}}$. We note that, from BP to BAs and to BSb, $F(\omega)$ doesn't change much because their sound velocities are similar (see Table 2 and the Supporting Information). Instead, the EPC strengths decrease significantly due to the decreased $m^*$ (see Table 1) and increased M. Therefore, the intravalley scattering rates decrease from BP to BAs and to BSb. Besides, the intravalley scatterings caused by the high-energy phonons are also decreased due to the decreased effective masses.

**CONCLUSION**

In conclusion, using the first-principles calculations and considering all the electron-phonon couplings, we have demonstrated that monolayer BX (X = P, As, Sb) with the honeycomb lattice, although they are polar and their band edges are located at $K$ point with multiple valleys, unusually have both high electron and hole mobility at room temperature (1200~14000 cm$^2$V$^{-1}$s$^{-1}$). Particularly, the intrinsic electron and hole mobility in monolayer of BSb achieve 8956.3 cm$^2$V$^{-1}$s$^{-1}$ and 14033.4 cm$^2$V$^{-1}$s$^{-1}$, respectively, which are the largest values among known 2D materials. Through our analysis, we conclude: (1) for the intravalley scatterings caused by low-energy acoustic phonons, high sound velocity which leads to small $F(\omega)$ and small effective mass as well as large atomic mass which leads to the small EPC, tend to result in small scattering rates; (2) for the intravalley scattering caused by high-energy optical phonons, small effective mass and large-energy optical phonons tend to result in small $F(\omega)$ and thus small scattering rates; (3) for the intervalley scatterings, out-of-plane orbitals and large momentum of optical phonon modes tend to result in the small EPC and thus small scattering rates. Our findings not only have demonstrated that 2D semiconductors can be comparable to 3D high-performance systems from the point view of carrier mobility, but also have provided useful insights for searching more high-mobility 2D systems.

**ASSOCIATED CONTENT**

**Supporting Information**

Supporting information for this article are available at XXXX.

Detailed calculation methods and analytical form of $F(\omega)$, analytical form of electron-phonon coupling, carrier scattering at the K valley for monolayer BP, carrier scattering at the K valley for monolayer BAs, and other supporting data.

Fig. S1. The phonon-energy resolved scattering rates, matching functions, and average electron-phonon coupling strength for hole and electron in monolayer BP.

Fig. S2. The phonon-branch resolved scattering rates in monolayer BP.

Fig. S3. The phonon-energy resolved scattering rates, matching functions, and average electron-phonon coupling strength for hole and electron in monolayer BAs.

Fig. S4. The phonon-branch resolved scattering rates in monolayer BAs.

Table S1. Scattering rates and mobilities for electrons in BP, BAs and BSb.

Table S2. Scattering rates and mobilities for holes in BP, BAs and BSb.

# AUTHOR INFORMATION

**Corresponding Author**

E-mail: [jhyang04@fudan.edu.cn](jhyang04@fudan.edu.cn)

**Author contributions**

J.-H. Y. conceived the idea, S.-R. S., S.-X. L., Y.-T. S., J.-H. Y. and X.-G. G. analyzed the data, and wrote the paper, S.-R. S. performed the calculations.

**Notes**

The authors declare that they have no competing interests.

# ACKNOWLEGEMENTS

This work was supported in part by the National Natural Science Foundation of China (Grant No. 12188101, 61904035, 11974078, 11991061) and the Shanghai Sailing Program(19YF1403100).

The supercomputing time was sponsored by the Fudan University High-End Computing Center.

# TABLES

Table 1. Calculated carrier effective masses and mobility in BP, BAs, BSb, and InN monolayers. Note that all the VBMs and CBMs locate at the K points, except that the CBM of InN locates at the Γ point. The hole and electron mobilities are given by the average numbers for the different directions. $m_e$ is the electron mass.

|     | Mass($m_e$) | | | | Mobility(cm$^2$/Vs) | |
| --- | --- | --- | --- | --- | --- | --- |
|     | hole | | electron | | hole | electron |
|     | armchair | zigzag | armchair | zigzag | | |
| BP  | 0.19 | 0.17 | 0.19 | 0.18 | 1607.8 | 1210.1 |
| BAs | 0.16 | 0.07 | 0.16 | 0.07 | 2675.3 | 2039.7 |
| BSb | 0.07 | 0.07 | 0.07 | 0.07 | 14033.4 | 8956.3 |
| InN | 1.29 | 1.48 | 0.08(Γ) | 0.08(Γ) | 8.8 | 366.5 |

Table 2. Calculated phonon group velocities or sound velocities for the in-plane acoustic branches in BP, BAs, BSb, InN and MoS$_2$ monolayers.

|     | Velocity(m/s) | | | |
| --- | --- | --- | --- | --- |
|     | TA | | LA | |
|     | armchair | zigzag | armchair | zigzag |
| BP  | 2392.4 | 2467.4 | 4048.7 | 4088.1 |
| BAs | 1614.4 | 1672.1 | 2743.2 | 2771.6 |
| BSb | 762.4 | 810.0 | 1990.4 | 1950.3 |
| InN | 1008.0 | 1086.7 | 2158.2 | 2113.0 |
| MoS$_2$ | 1230.3 | 1236.1 | 1994.4 | 1985.2 |

**FIGURES**

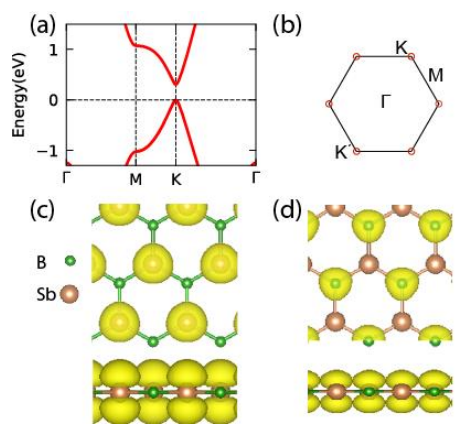

**Figure 1.** Band structures, band edge valleys and partial charge densities of the band edges in BSb monolayer. (a) Band structure of monolayer BSb. (b) The VBM (CBM) valleys in monolayer BSb. (c) Top and side view of the VBM partial charge densities in monolayer BSb. (d) Top and side view of the CBM partial charge densities in monolayer BSb. The B atom is represented by the small green ball, and the Sb atom is represented by the big brown ball.

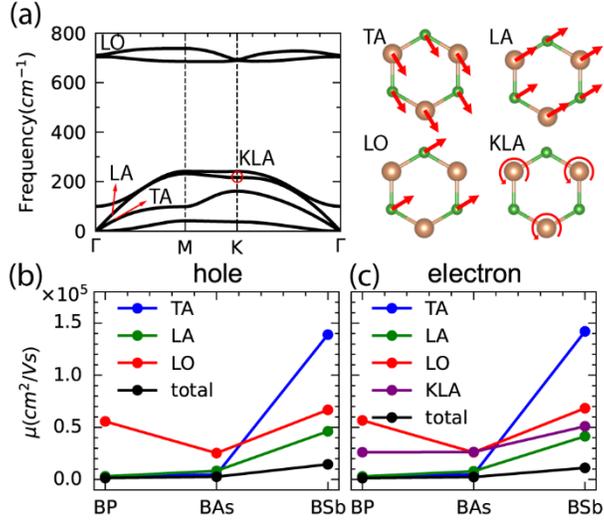

**Figure 2.** Phonon modes of BX and branch contributions to the total mobility. (a) Left: phonon band structure of BSb. The TA, LA, LO and KLA modes are indicated respectively; right: vibration modes of the transverse acoustic (TA), longitudinal acoustic (LA), longitudinal optical (LO) around the Γ point and phonon modes around the K point (KLA) that limit the carrier mobility. (b) The total hole mobility (black) and the TA (blue), LA (green), and LO (red) limited mobility for BP, BAs, BSb. (c) The total electron mobility (black) and the TA (blue), LA (green), LO (red), and KLA (purple) limited mobility for BP, BAs, BSb.

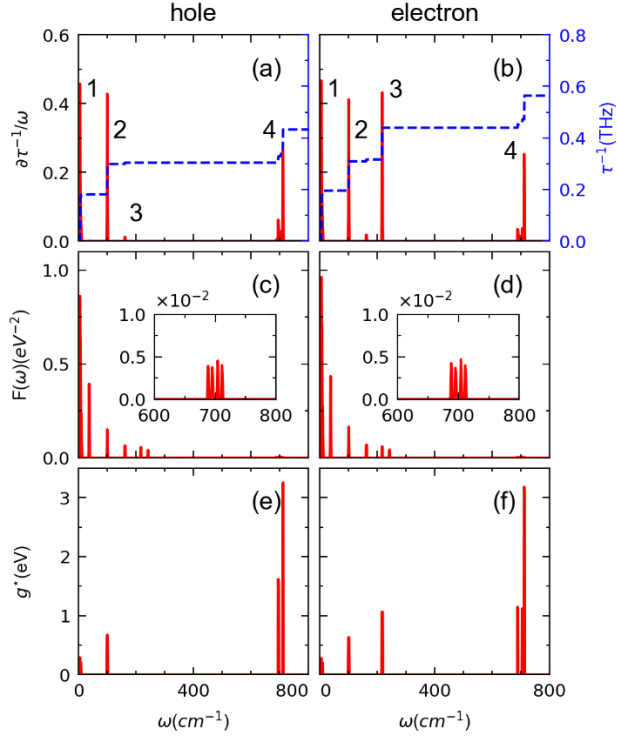

**Figure 3.** The phonon-energy resolved scattering rates, matching functions, and average electron-phonon coupling strength for hole and electron in monolayer BSb. (a) and (b) Phonon-spectra-decomposed hole and electron scattering rate, (c) and (d) phonon-energy resolved matching functions between the electronic band structure and phonon spectrum for hole and electron in monolayer BSb, (e) and (f) phonon-energy resolved average electron-phonon coupling strength for hole and electron in monolayer BSb. Note that, the blue lines in (a) and (b) indicate the integrated scattering rates.

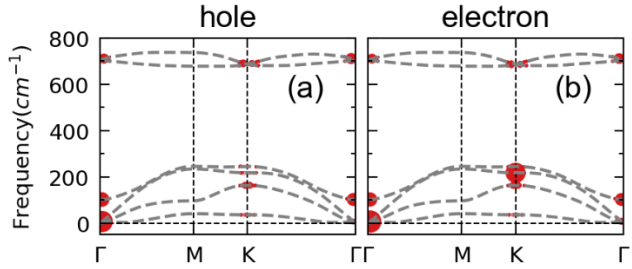

**Figure 4.** The phonon-branch resolved scattering rates in monolayer BSb. (a) The phonon-branch resolved scattering rates for the K-valley hole in BSb monolayer. (b) The phonon-branch resolved scattering rates for the K-valley electron in BSb monolayer. Note that, the sizes of the dots are proportional to the scattering rates assisted by the corresponding phonon modes. Note that, the contribution to the hole scattering rates around 200 cm$^{-1}$ is smaller than that for the electron case in monolayer BSb due to the smaller hole effective mass.

SUPPORTING INFORMATION for:

# Exceptionally high phonon-limited carrier mobility in BX (X = P, As, Sb) monolayers

Shiru Song,[†] Shixu Liu,[†] Yuting Sun,[†] Ji-Hui Yang[†‡]* and Xin-Gao Gong[†‡]

[†]Key Laboratory of Computational Physical Sciences (Ministry of Education), Institute of Computational Physics, Fudan University, Shanghai 200433, China
[‡]Shanghai Qi Zhi Institute, Shanghai 200230, China
Email: jhyang04@fudan.edu.cn

Detailed calculation methods and analytical form of $F(\omega)$, analytical form of electron-phonon coupling, carrier scattering at the *K* valley for monolayer BP, carrier scattering at the *K* valley for monolayer BAs, and other supporting data.

Fig. S1. The phonon-energy resolved scattering rates, matching functions, and average electron-phonon coupling strength for hole and electron in monolayer BP.

Fig. S2. The phonon-branch resolved scattering rates in monolayer BP.

Fig. S3. The phonon-energy resolved scattering rates, matching functions, and average electron-phonon coupling strength for hole and electron in monolayer BAs.

Fig. S4. The phonon-branch resolved scattering rates in monolayer BAs.

Table S1. Scattering rates and mobilities for electrons in BP, BAs and BSb.

Table S2. Scattering rates and mobilities for holes in BP, BAs and BSb.

**Supplementary Note S1: Computation Methods**

We perform first-principles calculations using the Quantum Espresso Package[1,2] with the norm-conserving pseudopotentials[3] and the Perdew-Burke-Ernzerhof (PBE) exchange-correlation functional.[4] The kinetic energy cutoffs for wave functions and charge density are set to 80 and 320 Ry, respectively. The atomic coordinates for monolayer BP, BAs, BSb, InN are optimized using a 18 × 18 × 1 k mesh until the force acting on each atom becomes less than 0.0001 Ry/Bohr. Monolayer BXs are described using a vacuum-slab model. The length of the cells in the out-of-plane direction is 20 Å. The spin-orbit coupling effect is considered for BSb in the calculations. Calculations of electron-phonon couplings and carrier mobilities are performed using the EPW code.[5,6] The electron-phonon matrix elements are initially computed on a 9 × 9 × 1 electric grid and a 9 × 9 × 1 phonon grid, using density functional perturbation theory[7] (DFPT). Then the electron-phonon matrix elements are subsequently interpolated onto fine grids, 300 × 300 × 1 electronic grid and a 300 × 300 × 1 phonon grid, using maximally localized Wannier functions.[8,9] The carrier concentration for computing the mobilities is set to $10^{16}$ cm$^{-3}$, and the temperature is set to 300 K for all systems.

**Supplementary Note S2: Analytical form of $F(\omega)$.**

1. Acoustic phonons
   The energy band for the low-field mobility is described by a simple parabolic band: $\varepsilon_k = \frac{\hbar^2 k^2}{2m^*}$, where $m^*$ is the effective electron mass, **k** is measured with respect to the $K, K'$ points in the Brillouin zone. In the long-wavelength limit, the frequency of the transverse acoustic (TA) and longitudinal acoustic (LA) modes are given by linear relation: $\omega_{q\lambda} = v_\lambda q$, where $v_\lambda$ is the in-plane sound velocity. Except at very low temperatures, $\hbar\omega \ll k_B T$ implies $n_q \sim k_B T/\hbar\omega_q \gg 1$ for the Bose-Einstein distribution.
   For a conduction band and a branch of acoustic phonon,

$$F(\omega) = \frac{1}{\hbar} \int \frac{d^2q}{\Omega_{BZ}} [n_{q\nu} \delta(\varepsilon_{nk} - \varepsilon_{mk+q} + \hbar\omega_{q\nu})] \delta(\omega - \omega_{q\nu})$$

$$= \frac{1}{\hbar} \int_D \frac{d^2q}{\Omega_{BZ}} \left[\frac{k_B T}{\hbar v_\lambda q} \delta\left(\hbar v_\lambda q - \frac{\hbar^2 q^2}{2m^*}\right)\right] \delta(\omega - v_\lambda q)$$

$$= \frac{2\pi}{\hbar} \int_0^{q_D} \frac{q dq}{\Omega_{BZ}} \left[\frac{k_B T}{\hbar v_\lambda q} \frac{1}{\hbar v_\lambda} \delta\left(q - \frac{2m^* v_\lambda}{\hbar}\right)\right] \frac{1}{v_\lambda} \delta\left(q - \frac{\omega}{v_\lambda}\right)$$

$$= \frac{2\pi k_B T}{\Omega_{BZ} \hbar^3 v_\lambda^3} \delta\left(\frac{\omega}{v_\lambda} - \frac{2m^* v_\lambda}{\hbar}\right) = \frac{2\pi k_B T}{\Omega_{BZ} \hbar^3 v_\lambda^2} \delta\left(\omega - \frac{2m^* v_\lambda^2}{\hbar}\right)$$

(Because the intersection line between the electronic band and the phonon band is near the Γ point, the interval of integration is set to a disk D of radius $q_D$, which contains the intersection line.)

2. Optical phonons

   For a conduction band and a flat optical phonon branch with a constant frequency of $\omega_o$,

$$F(\omega) = \frac{1}{\hbar} \int \frac{d^2q}{\Omega_{BZ}} [(f^0_{mk+q} + n_{q\nu}) \delta(\varepsilon_{nk} - \varepsilon_{mk+q} + \hbar\omega_o)] \delta(\omega - \omega_o)$$

$$= \frac{1}{\hbar} (f^0_{mk+q} + n_{q\nu})|_{\varepsilon = \hbar\omega_o} \int_D \frac{d^2q}{\Omega_{BZ}} \left[\delta\left(\hbar\omega_o - \frac{\hbar^2 q^2}{2m^*}\right)\right] \delta(\omega - \omega_o)$$

$$= \frac{2\pi}{\hbar} (f^0_{mk+q} + n_{q\nu})|_{\varepsilon = \hbar\omega_o} \int_0^{q_D} \frac{q dq}{\Omega_{BZ}} \left[\delta\left(\hbar\omega_o - \frac{\hbar^2 q^2}{2m^*}\right)\right] \delta(\omega - \omega_o)$$

$$= \frac{2\pi m^*}{\Omega_{BZ} \hbar^3} (f^0_{mk+q} + n_{q\nu})|_{\varepsilon = \hbar\omega_o} \delta(\omega - \omega_o)$$

$$\propto m^* \left[\frac{1}{\exp\left(\frac{\hbar\omega_o}{k_B T}\right) + 1} + \frac{1}{\exp\left(\frac{\hbar\omega_o}{k_B T}\right) - 1}\right] \delta(\omega - \omega_o)$$

**Supplementary Note S3: Analytical form of electron-phonon coupling.**

The electron-phonon coupling for the phonon mode with wave vector **q** and branch $\lambda$ is given by[10]

$$g^\lambda_{kq} = \sqrt{\frac{\hbar}{2MN\omega_{q\lambda}}} M^\lambda_{kq},$$

where $\omega_{q\lambda}$ is the phonon frequency, $M$ is effective mass, $N$ is the number of unit cell in the crystal. The coupling matrix element is defined by

$$M^\lambda_{kq} = \langle k + q | \delta V_{q\lambda}(r) | k \rangle,$$

where **k** is the wave vector of the carrier, $\delta V_{q\lambda}(r)$ is the change of the effective potential corresponding to vibrational normal mode **q**, $\lambda$. The carrier is scattered from the wave vector **k** to **k + q**, by the change of the effective potential $\delta V_{q\lambda}(r)$.

For the acoustic phonons, the coupling matrix element in the long-wavelength limit is given by

$$M^\lambda_{kq} = \Xi_\lambda q,$$

where $\Xi_\lambda$ is the acoustic deformation potential.

For scattering caused by the acoustic phonons, the scattering rate for the electron

$$\tau_{nk}^{-1} = \frac{2\pi}{\hbar} \int \frac{d^2q}{\Omega_{BZ}} \frac{\hbar}{2MN\omega_{q\lambda}} (\Xi_\lambda q)^2 [(f_{mk+q}^0 + n_{qv})\delta(\varepsilon_{nk} - \varepsilon_{mk+q} + \hbar\omega_{qv})]$$

$$= \frac{(2\pi)^2}{\hbar} \int \frac{qdq}{\Omega_{BZ}} \frac{\hbar}{2MNv_\lambda q} (\Xi_\lambda q)^2 \left[\frac{k_B T}{\hbar v_\lambda q} \delta\left(\hbar v_\lambda q - \frac{\hbar^2 q^2}{2m^*}\right)\right]$$

$$= \frac{2\pi^2 k_B T \Xi_\lambda^2}{\hbar MN v_\lambda^2} \int \frac{qdq}{\Omega_{BZ}} \delta\left(\hbar v_\lambda q - \frac{\hbar^2 q^2}{2m^*}\right) = \frac{(2\pi)^2 k_B T \Xi_\lambda^2 m^*}{\hbar^3 MN v_\lambda^2 \Omega_{BZ}} = \frac{k_B T \Xi_\lambda^2 m^*}{\hbar^3 \rho v_\lambda^2},$$

where $\rho$ is the atomic mass density per area.

The derivative of scattering rate with respect to $\omega$,

$$\frac{\partial \tau_{nk}^{-1}}{\partial \omega} = \frac{2\pi}{\hbar} \sum_{m,v} \int \frac{d^2q}{\Omega_{BZ}} \frac{\hbar}{2MN\omega_{q\lambda}} (\Xi_\lambda q)^2 [(f_{mk+q}^0 + n_{qv})\delta(\varepsilon_{nk} - \varepsilon_{mk+q} + \hbar\omega_{qv})]\delta(\omega - \omega_{qv})$$

$$= \frac{(2\pi)^2}{\hbar} \int \frac{qdq}{\Omega_{BZ}} \frac{\hbar}{2MNv_\lambda q} (\Xi_\lambda q)^2 \left[\frac{k_B T}{\hbar v_\lambda q} \delta\left(\hbar v_\lambda q - \frac{\hbar^2 q^2}{2m^*}\right)\right]\delta(\omega - v_\lambda q)$$

$$= \frac{2\pi^2 k_B T \Xi_\lambda^2}{\hbar MN v_\lambda^2} \int \frac{qdq}{\Omega_{BZ}} \delta\left(\hbar v_\lambda q - \frac{\hbar^2 q^2}{2m^*}\right)\delta(\omega - v_\lambda q)$$

$$= \frac{(2\pi)^2 k_B T \Xi_\lambda^2 m^*}{\hbar^3 MN v_\lambda^3 \Omega_{BZ}} \delta\left(\frac{\omega}{v_\lambda} - \frac{2m^* v_\lambda}{\hbar}\right) = \frac{(2\pi)^2 k_B T \Xi_\lambda^2 m^*}{\hbar^3 MN v_\lambda^2 \Omega_{BZ}} \delta\left(\omega - \frac{2m^* v_\lambda^2}{\hbar}\right),$$

For the average electron-phonon coupling strength $g_{nk}^*(\omega)$,

$$|g_{nk}^*(\omega)|^2 = \frac{\frac{1}{2\pi}\left(\frac{\partial \tau_{nk}^{-1}}{\partial \omega}\right)}{F(\omega)} = \frac{1}{2\pi} \frac{(2\pi)^2 k_B T \Xi_\lambda^2 m^*}{\hbar^3 MN v_\lambda^2 \Omega_{BZ}} \frac{\Omega_{BZ} \hbar^3 v_\lambda^2}{2\pi k_B T} = \frac{\Xi_\lambda^2 m^*}{MN}$$

**Supplementary Note S4: The carrier scattering at the *K* valley for monolayer BP.**

As seen in Figs. S1a and S1b, the electron and hole scattering are very similar to each other and the carrier scattering rates are mainly contributed by the low-energy phonons (0–100 cm$^{-1}$ or 0–12 meV) corresponding to the intravalley scattering around the band edges, as seen from the phonon-branch-resolved scattering rates in Figs. S2a and S2b showing that the phonons involved in the scattering process mainly have small momentum. A small peak, which is contributed by phonons around 950 cm$^{-1}$, comes from the intravalley scattering (see Figs. S2a and S2b). For electron, an additional peak around 500 cm$^{-1}$ comes from the intervalley scattering, as the contributions are mainly from phonons with momentum **q**=**K** (see Fig. S2b). There is no peak around 500 cm$^{-1}$ for hole in Fig. S1a, as small electron-phonon interaction at that region, as seen in Fig. S1c.

**Supplementary Note S5: The carrier scattering at the *K* valley for monolayer BAs.**

As seen in Figs. S3a, the electron scattering for BAs is mainly contributed by the low-energy phonons (0–100 cm$^{-1}$ or 0–12 meV) corresponding to the intravalley scattering around the band edges, as seen from the phonon-branch-resolved scattering rates in Figs. S4a showing that the phonons involved in the scattering process mainly have small momentum. Two small peaks are contributed by phonons around 800 cm$^{-1}$(see Figs. S3a). The lower one comes from the intervalley scattering, while the higher one comes from the intravalley scattering (see Figs. S4a). For electron, similar scattering peaks located around 0 cm$^{-1}$, 200 cm$^{-1}$ and 800 cm$^{-1}$. Besides, there is an additional peak around 300 cm$^{-1}$ comes from the intervalley scattering, as the contributions are mainly from phonons with momentum **q=K** (see Fig. S4b). There is no peak around 300 cm$^{-1}$ for hole in Fig. S3a, as small electron-phonon interaction at that region, as seen in Fig. S3c.

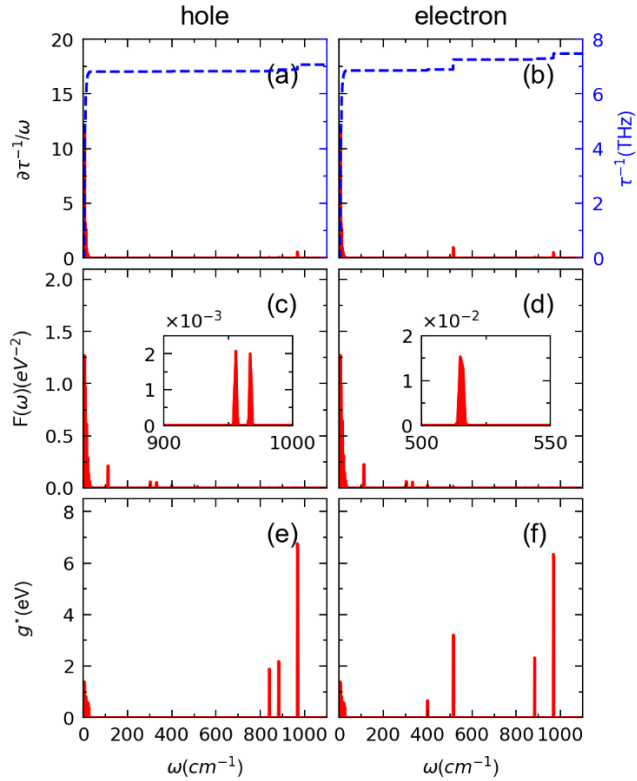

**Figure S1.** The phonon-energy resolved scattering rates, matching functions, and average electron-phonon coupling strength for hole and electron in monolayer BP. (a) and (b) The phonon-spectra-decomposed hole and electron scattering rate, (c) and (d) phonon-energy resolved matching functions between the electronic band structure and phonon spectrum for hole and electron in monolayer BP, the inset in (c) and (d) indicate the non-zero valued matching functions

at the position of the peak in the (a) and (b), respectively. (e) and (f) phonon-energy resolved average electron-phonon coupling strength for hole and electron in monolayer BP. The insets in (e) and (f) indicate the non-zero valued average electron-phonon coupling strength for hole and electron between 500 cm$^{-1}$ and 550 cm$^{-1}$, which contributes the peaks in (a) and (b), respectively. Note that, the blue lines in (a) and (b) indicate the integrated scattering rates.

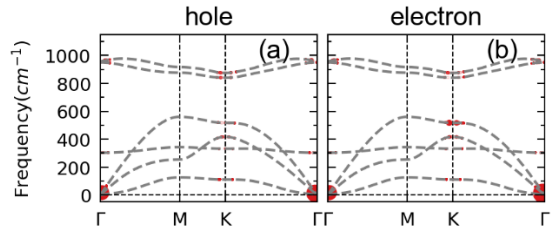

**Figure S2.** The phonon-branch resolved scattering rates in monolayer BP. (a) The phonon-branch resolved scattering rates for the $K$-valley hole in BP monolayer. (b) The phonon-branch resolved scattering rates for the $K$-valley electron in BP monolayer. Note that, the sizes of the dots are proportional to the scattering rates assisted by the corresponding phonon modes. The contribution to the hole scattering rates around 500 cm$^{-1}$ is near zero, while non-zero for the electron, which correspond to the inter-valley scattering.

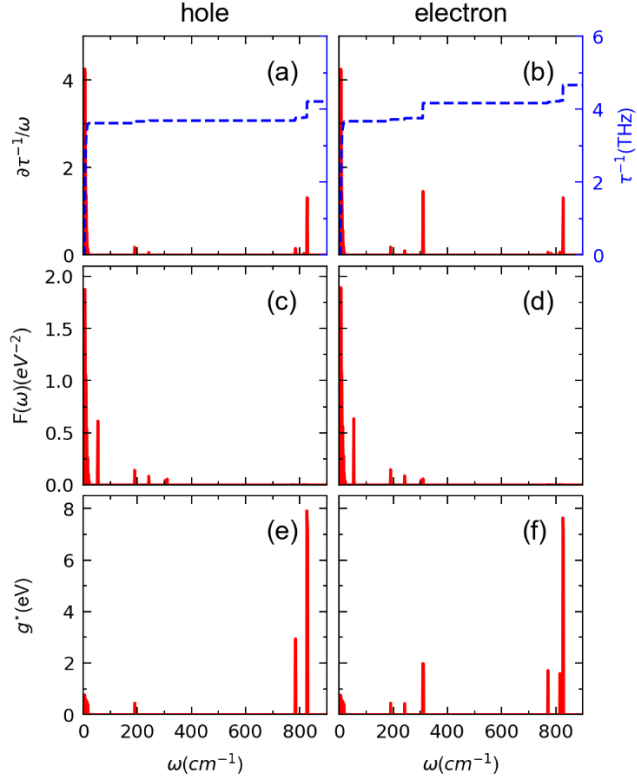

**Figure S3.** The phonon-energy resolved scattering rates, matching functions, and average electron-phonon coupling strength for hole and electron in monolayer BAs. (a) and (b) The phonon-spectra-decomposed hole and electron scattering rate, (c) and (d) phonon-energy resolved matching functions between the electronic band structure and phonon spectrum for hole and electron in monolayer BAs, the inset in (c) and (d) indicate the non-zero valued matching functions at the position of the peak in the (a) and (b), respectively. (e) and (f) phonon-energy resolved average electron-phonon coupling strength for hole and electron in monolayer Bas, the inset in (e) and (f) indicate the non-zero valued average electron-phonon coupling strength at the position of the peak in the (a) and (b), respectively. Note that, the blue lines in (a) and (b) indicate the integrated scattering rates.

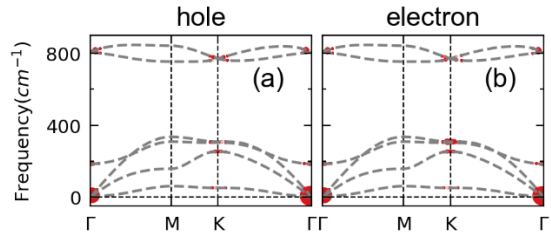

**Figure S4.** The phonon-branch resolved scattering rates in monolayer BAs. (a) The phonon-branch resolved scattering rates for the $K$-valley hole in BAs monolayer. (b) The phonon-branch resolved scattering rates for the $K$-valley electron in BAs monolayer. Note that, the sizes of the dots are proportional to the scattering rates assisted by the corresponding phonon modes. The contribution

to the hole scattering rates around 300 cm$^{-1}$ is smaller than that for the electron case in monolayer BAs.

Table S1. Scattering rates and mobilities for electrons in BP, BAs and BSb.

| Electron | BP | | BAs | | BSb | |
|---|---|---|---|---|---|---|
| | $\tau^{-1}$(THz) | $\mu$(cm$^2$/Vs) | $\tau^{-1}$(THz) | $\mu$(cm$^2$/Vs) | $\tau^{-1}$(THz) | $\mu$(cm$^2$/Vs) |
| TA | 3.58 | 2658.0 | 2.27 | 4836.3 | 0.04 | 142067.2 |
| LA | 3.28 | 2898.6 | 1.40 | 7877.6 | 0.15 | 41440.2 |
| KLA | 0.36 | 26128.5 | 0.42 | 26373.2 | 0.12 | 51000.7 |
| ZO | 0 | | 0.05 | 212623.3 | 0.11 | 55497.6 |
| LO | 0.17 | 56708.2 | 0.42 | 26052.6 | 0.09 | 68355.2 |
| Total | 7.47 | 1272.0 | 4.66 | 2359.3 | 0.56 | 11140.6 |

The mobilities in the table is calculated using the scattering rates of band edges, according to the formulation: $\mu = e\frac{\tau}{m^*}$.

Table S2. Scattering rates and mobilities for holes in BP, BAs and BSb.

| Hole | BP | | BAs | | BSb | |
|---|---|---|---|---|---|---|
| | $\tau^{-1}$(THz) | $\mu$(cm$^2$/Vs) | $\tau^{-1}$(THz) | $\mu$(cm$^2$/Vs) | $\tau^{-1}$(THz) | $\mu$(cm$^2$/Vs) |
| TA | 3.57 | 2738.0 | 2.29 | 4807.4 | 0.05 | 138986.6 |
| LA | 3.26 | 2999.2 | 1.33 | 8275.4 | 0.14 | 46187.5 |
| KLA | 0 | | 0 | | 0 | |
| ZO | 0 | | 0.05 | 220293.1 | 0.12 | 53389.1 |
| LO | 0.18 | 55686.0 | 0.43 | 25302.4 | 0.09 | 66781.8 |
| Total | 7.07 | 1382.4 | 4.21 | 2611.1 | 0.43 | 14508.9 |